\documentclass[aps,prl,twocolumn,groupedaddress, showpacs]{revtex4-1}
\usepackage{graphicx}

\begin{document}

\title{A Topological Phase Transition in the Scheidegger Model of River Networks}
\author{Jacob N. Oppenheim}
\email[]{joppenheim@rockefeller.edu}
\author{Marcelo O. Magnasco}
\affiliation{Laboratory of Mathematical Physics, Rockefeller University, New York, New York 10065}

\date{\today} %Change this upon Submission

\begin{abstract}
Transport networks are found at the heart of myriad natural systems, yet are poorly understood, except for the case of river networks.  The Scheidegger model, in which rivers are convergent random walks, has been studied only in the case of flat topography, ignoring the variety of curved geometries found in nature.  Embedding this model on a cone, we find a convergent and a divergent phase, corresponding to few, long basins and many, short basins, respectively, separated by a singularity, indicating a phase transition.  Quantifying basin shape using Hack�s Law $l\sim a^h$ gives distinct values for $h$, providing a method of testing our hypotheses.  The generality of our model suggests implications for vascular morphology, in particular differing number and shapes of arterial and venous trees.
\end{abstract}


%\maketitle must follow title, authors, abstract, \pacs, and \keywords
\maketitle

\section{Introduction}
Transport networks occur in a vast variety of natural systems.  From river basins to leaf venation, we observe a spectrum of solutions to the problem of optimal distribution through a landscape, given a set of constraints on the construction of links.  Previous work on treelike networks originates in the study of river basins, which can be modeled as convergent random walks, or equivalently as flow through a linear slope field with additive noise.  This approach, the Scheidegger Model \cite{ScheiPaper}, does not incorporate the effects of convergent and divergent topologies, which occur both in geomorphology (e.g. endorheic basins) and in biologically relevant systems.  The vasculature of the retina, for instance, may be viewed as divergent for the arteries, which originate at the center, and convergent for the veins, which terminate there.  Another example of this morphology occurs in the functional units of organs, such as the liver, have distinct zones where the arteries originate and the veins terminate, respectively.  Convergence and divergence in river networks originates in the presence of curvature in the underlying topography.  Thus, to begin to understand the morphologies of transport networks found in these biologically-relevant examples, we investigate the effects of curving the manifold in which we embed the Scheidegger Model, from a plane into a cone \cite{BohnMag,RodriguezIturbe}.

The study of river networks depends upon the ability to impose a hierarchy on the streams, dividing the main channel from its tributaries.  Ordering by magnitude, where each node is labeled with the number of drains, becomes rapidly intractable, due to the exponential growth of binary trees.   Strahler stream ordering assigns all tips a value of one.  At every junction, the resulting stream takes the value of the larger of its two inputs, unless they are the same, in which case it is augmented by one \cite{Strahler}.  Binary trees grow exponentially (the number of nodes within a distance $r$ of the center goes as $2^r$) and cannot be embedded linearly into Euclidean space, which grows only algerbraically (the volume of space goes as $r^d$, where $d$ is the number of spatial dimensions).  Strahler stream ordering, which requires pairs of order $n-1$ streams to create one of order $n$, is inherently logarithmic and able to ``fit" a binary tree into Euclidean space \cite{DoddsRothman}.  Horton first examined the scaling of the length of, number of and area drained by a link in a river network with its Strahler order.  His work and subsequent examinations of examples around the world showed that these three quantities nearly always scale in exponential fashion, defining an empirical range of length, bifurcation, and area ratios respectively \cite{Horton, MaritanPRE}.  These scaling relationships originate in the logarithmic nature of Strahler stream ordering.  The addition of curvature, which changes the scaling of area with radius from $a\sim r^2$, affects the ability to accommodate a binary tree.  Spaces with negative intrinsic curvature grow exponentially and have no need for logarithmic embedding.  There is thus an underlying connection between Horton's scaling laws and curvature of the embedding topology.

Numerous other scaling relationships (usually power law) have been observed in river networks(see table in \cite{DoddsRothman}).  Notably, it has also been found that $l\sim a^h$, with $l$ the mainstream length and $a$ the basin area, a relationship known as Hack' law  \cite{DoddsRothman}.  By relating basin morphology to a correlate of flow volume (the area drained), Hack's Law provides a quantification of basin shape, useful for vascular and other distribution trees\cite{RodriguezIturbe, DoddsRothman}.

Scheidegger (1967) proposed a simplified model of river networks, in order to better understand the origin of these scaling laws.  Consider a hexagonal grid of points, tilted out of the plane.  Each point drains downhill and moves with equally probability to either the right or the left.  Basins are collections of convergent random walks.  The Hack exponent is 2/3: a basin is the area between two random walks (the neighboring streams).  This distance is itself a random walk; the area between two such random walks should scale as $l^{3/2}$ \cite{ScheiPaper, Takayasu, Dhar}.  The Scheidegger model may also be created by adding a slope field plus random noise to a lattice and requiring downhill flow from every point \cite{ScheiBook, DoddsRothman} .  This form is both a simplification of the physics of streamflow and amenable to embedding in convergent and divergent topologies.  Adding a radial instead of a unidirectional slope embeds the model in a cone \footnote{While all the intrinsic curvature in a cone is found in the tip, its effects are felt throughout the cone}.  A negative radial slope models the (temporary) divergent flow off of a mountaintop\footnote{Permanent channelization only occurs where the Laplacian of the slope field is negative; while divergent river networks occur, they are fugitive}, while a positive value represents the flow into an endorheic basin or a deep valley.  The effects of curvature have been investigated previously for their effects on channelization and stream head formation, but not for their effects on network morphology \cite{ScheiBook, smith1972stability, izumi2000linear}.

\section{Simulation Procedure}
Our procedure was to generate a grid of $5r_0^2$  random points in an annular region of outer radius $r_0$ and inner radius $r_0/5$ with connectivities defined through the Delaunay triangulation.  To eliminate edge effects, points were placed in the square region of width $2r_0$ centered on the origin and then removed if they fell outside the desired annulus. Each point was assigned a height equal to its Euclidean distance from the origin times the radial slope of the simulation, $m$, plus frozen white noise, $\eta$, giving: $z= mr+\eta$.  

Each point was then connected to its neighbor with the lowest value.  If all neighbors had a larger z value, then the point was assumed to be the mouth of a river network.  These points almost always only occurred on the edge of the region, except in the flattest cases, see below.  From the resulting connectivities matrix, we could recursively calculate the magnitude and Strahler order numbers at each point in the grid.  Additionally, all points that flowed into a single mouth were collected as a basin allowing us to collect volume, length, and order statistics.

The following results are robust to lattice geometry. Unless otherwise noted, all simulations were conducted with the following parameters: a radius of 67 units, a radial slope of -1 or 1, and a noise strength of 0.1.  

\section{Results and Discussion}
\begin{figure}
\includegraphics{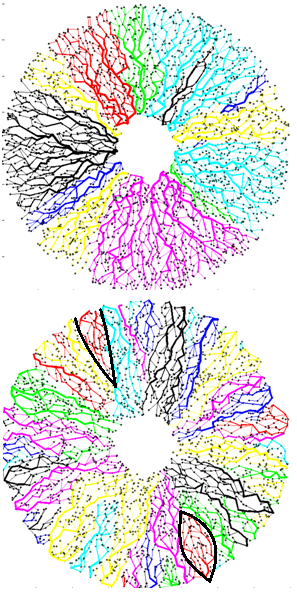}
\caption{(Color Online) Simulated Networks for convergent ($m=1$) and divergent ($m=-1$)  embedding topologies on the top and bottom, respectively.  Basins, defined as collections of points with the same outlet, are colored the same.  Note the differences in shape and size of basins.  In the convergent case, most basins take the form of large sectors of the annular region, whereas in the divergent case there are basins at all length scales, with a characteristic leaflike shape (e.g. the heavy-outlined basins).\label{Networks}}
\end{figure}

At positive and negative values of the radial slope, m, two distinct phases emerge.   In the case of a convergent network, $m>0$, there are relatively few, but large basins, whereas for a divergent network,$ m<0$, there are many more, considerably smaller and narrower basins.  See Figure \ref{Networks}, where each basin is represented by a single color.  The basins in the divergent case all appear to share the same characteristic, leaflike shape: narrow at the top, widening in the center, and tapering towards the mouth, reminiscent of tilings of the Poincare disk \cite{Coxeter, CircleLimits}.

\begin{figure}
\includegraphics{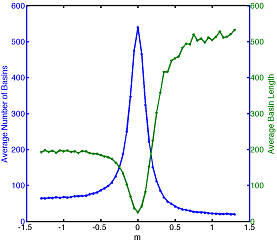}
\caption{(Color Online) Phase Transition.  As $m\rightarrow 0$  from both sides, the number of basins approaches a singularity and the average of basin length falls to zero, both in power law fashion, with exponents respectively $=-0.96\pm0.05$ and $=0.94\pm0.07$.  Results here come from averaging 40 simulations at $r=67$. \label{PhaseTrans}}
\end{figure}

Iterating simulations from $m=-1.5$ to $m=1.5$, to investigate the crossover region, we found a sharp increase in the average number of basins as we approached m=0 with the power-law scaling indicative of a singularity, $N_{basins}\sim m^\gamma$, (see figure \ref{PhaseTrans}).  Additional properties such as the average basin length exhibit a zero, $\bar{l}\sim m^\delta$, when m vanishes, again with power law scaling.  These results lead us to conclude that there is a phase transition at zero radial slope for our modified Scheidegger Model.
We used finite-size scaling analysis to determine the critical exponent(s).  Conducting simulations at exponentially distributed network radii from 4 to 100, we determined a value of 1 for the finite-size scaling exponent, as predicted by the conventional theory of finite-size effects  \cite{Goldenfeld}.  From these results, we found a value of $=-0.96\pm0.05$  for $\gamma$, the critical exponent for $N_{basins}$.  Extrapolating to the thermodynamic limit, we can estimate $N_{basins}$ in the divergent case as $3r/2$ and in the convergent case, $r/2$. Hence, the ratio of basins to lattice points vanishes.  

\begin{figure}
\includegraphics{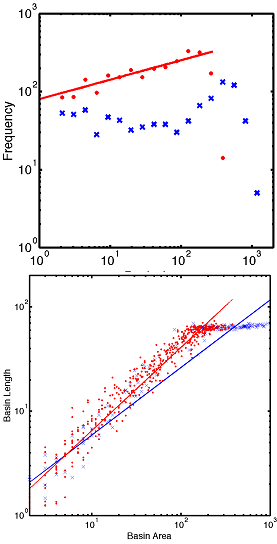}
\caption{(Color Online) Properties of convergent (blue) and divergent Basins (red).  On the top, the frequency of basin sizes is plotted logarithmically.  The convergent basins, in blue (dark gray) display an enhancement at large sizes, whereas there is a scale free distribution of basins in the divergent case, in red (light gray).  On the bottom, the Hack's law relationship is plotted, $\ln a $ vs $ \ln l$; data points are x's and a linear regression is overlaid.  Note the differing slopes for the convergent case it is approximately 2/3, the Scheidegger value.  For the divergent case it is nearly unity.\label{Hack}}
\end{figure}

Beyond the average number and size of a basin, we examine the distribution of basin sizes and shapes.  We took the cases $m=\pm1$ at a simulation radius of 67 and aggregated 40 simulation runs.  Given the wide dispersal in basin sizes, we binned the data logarithmically and examined the logarithm of the number of counts.  In the convergent case, there was a sharp enhancement of basins at the largest size, but little other structure.  In the divergent case, the data fell onto a line of slope $\approx0.25$, indicating a scale-free distribution with exponent $\approx-0.75$ (see Fig \ref{Hack}, upper panel), quantifying the self similarity of basin shapes seen in Figure \ref{Networks}.  Examining $\ln a$ vs. $\ln l$, in the convergent case, we observe a knee in the distribution, as finite size effects greatly limited mainstream length, see Figure \ref{Hack}, lower panel.  The largest basins were three decades greater in area and two in length than the smallest, matching the Scheidegger prediction of $h\approx2/3$.  In the divergent case, finite-size effects were generally much less important.  We found a (nearly) linear relationship between $a$ and $l (h\approx1)$.  Since basins on a divergent cone tend to be narrow and elongated with little branching, this value is not unexpected.  The Hack exponent provides a useful, and easily calculated, metric for the differentiation of convergent and divergent basins.

The maximum Strahler order observed in networks we could reliably simulate, was found to be 6 and to only occur in convergent topologies.  With such a low maximum order, it is impossible to properly define and calculate the bifurcation and other Hortonian ratios  \cite{DoddsRothman}.  We examined instead the distribution of Strahler numbers and found no difference in the PDF of Strahler numbers for 5th and 6th order basins between divergent and convergent networks.  While the size and shape of basins may vary wildly due to the embedding manifold, the hierarchy of the links within each basin does not change.
	
\section{Conclusions and Applications}
	When one embeds the Scheidegger model in a curved manifold, it is not the stream ordering properties that change, but rather the way that the manifold is tiled by the river basins.  We have observed severe changes in the number and length of basins and in the Hack exponent, but not the distribution of Strahler orders.  The same within-basin ordering principles yield grossly different bulk morphologies at the level of basins.  In essence, the curving of the embedding manifold affects the shapes of the basins themselves, while the area within each basin may be treated as locally flat.

We observe two distinct phases that can only be transformed into one another after passage through a singularity, where the average number of basins blows up to order $N$ from order $\sqrt{N}$ in convergent and divergent embedding topologies.  Given the minimal specifications of our implementation of the Scheidegger model, a noisy set of heights, z, and flow between neighbors chosen by $\Delta z$, we may expect that the general features are reflected in natural systems.  Specifically, we predict that convergent and divergent networks will exhibit different morphologies, quantified by the Hack exponent. 

A potential system for testing our model is vascular networks.  As mentioned previously, divergent topography does not lead to permanent channelization, preventing us from comparing the flow off of mountaintops to that into endorheic basins.  Two-dimensional vascular networks are typified by leaf venation; that system, however, does not exhibit a difference between convergence and divergence: flow both to and from the stem is carried in the same veins.  Nearly two-dimensional systems may be found in certain tissues, such as the retina. At the smallest level, however, vascular networks become loopy meshes.  At larger scales, these networks are treelike.  We hypothesize a morphological difference between arterial and venous trees.  Assuming a uniform drainage density, here a constant level of blood demand throughout thetissue in question, we may compare the length of these trees with the cross-sectional area of the terminal arteries and veins, as it correlates with flow volume \cite{Gray, Murray1, Murray2}.  

Given a large enough sample of arteries and veins, the distribution of ``basin" sizes, which in vasculature would correspond with vessel cross-sectional area, could be investigated.  We predict a scale-free distribution of terminal artery sizes and a distribution peaked at the system size for veins.  Given the simplicity and generality of the assumptions of our model, the absence of these properties would warrant careful study of the mechanisms shaping the growth and form of vasculature.

While this system does not pose a perfect test of our model,  as it exists in three dimensions, and at its lowest level (the capillaries), it is full of loops, naive models of angiogenesis, in which vascularature growth is dependent on the concentration of VEGF, or other chemical signals, map directly onto our version of the Scheidegger model \cite{GlazierVasc, GlazierVascTumor}. The concentration of a signaling molecule originates in a certain point (or area), providing a radial slope due to diffusion, and Poisson noise provides the the fluctuations to impart randomness into the gradient.  

While nearly all vascular networks exist in three dimensions, we do not expect the dramatic differences between convergent and divergent networks to disappear with the addition of another Euclidean dimension.  As binary trees grow exponentially, the difference between two and three dimensions that grow algerbraically should be immaterial.  The differences between convergence, where streams are forced together, and divergence, where the additional space would amplify the scale-free distribution of basin sizes should be amplified.  At vanishing $m$, $N_{basins}$ will still be of order $r^2$, preserving the phase transition.  A proper understanding of distribution networks, then, must account for the severe differences between convergence and divergence, or explain their absence, given that they arise from the most basic theoretical assumptions.

\begin{acknowledgments}
JO and MM would like to thank Eleni Katifori, Alex Petroff, and Carl Modes for helpful discussions throughout.  Supported in part by the National Science Foundation under grant NSF PHY-1058899
\end{acknowledgments}

\section{Appendix}
\appendix
\section{Additional Materials and Methods}
Simulations were conducted in MATLAB with general procedure as follows: a set of $5r_0^2$ random points were generated in the region $[-r_0,r_0]$x$[-r_0,r_0]$, with $r_0$ equal to 67 unless otherwise noted (hence 89780 points were initially generated).  All points outside the circle $x^2+y^2=r_0^2$ were then discarded.  The remaining points had nearest neighbors defined by the Delaunay triangulation.  Each point was then assigned a height equal to $z=mr+\eta$, with $\eta$ being frozen white noise, whose strength could be adjusted in each simulation.  

Each point was then connected to the lowest of its nearest neighbors (minimum $z$ value).  Those points that did not have a neighbor with a lower $z$ value were presumed to be basin mouths (or lakes, see below).  After generating the connectivity matrix, we could calculate all of the necessary network quantities.  Magnitude and Strahler number were calculated recursively: we found the tips of all streams by searching the connectivity matrix for points with no inflow.  These were assigned a value of one.  We then took their mouths and ran the same code, augmenting the magnitude number by one, and keeping the Strahler number the same.  At all merges, the magnitude numbers were added and the Strahler numbers were augmented according to the rule: $c=\sup(a,b)$ or $c=a+1$ if $a=b$. 

Basins were identified by a recursive algorithm acting in the opposite direction: starting with the mouths, defined as points with no outflows.  All points flowing into each mouth were found form the connectivity matrix; the find inflow algorithm was then run on these points.  Lists of the point addresses within each basin could be converted into mainstream lengths, by calculating the Euclidean distance between mouth and furthest stream tip, and basin lengths, by summing the number of points.  The basin area came directly from the magnitude number at the mouth, as a uniform drainage density was assumed.

We performed these simulations over hundreds of trials at values of $m$ from -1.5 to 1.5, usually sampling every 0.05.  Data on the Hack's Law relationship and the distribution of basin sizes came from simulations at $r=67$.  For all other simulations we sampled at logarithmically-distributed basin sizes from $r=4$ to $r=100$ $(4,6,9,13,20,30,45,67,100)$.

Preliminary simulations on square and hexagonal grids were also conducted. However, the degree of noise necessary to observe a difference in behavior with changing curvature made simulating nearly flat topographies extremely difficult, as the number of, ``lakes" or basins that terminated not at the edge of the annular region was improperly large.  To remove a lake, one must simulate filling by raising the z-value at a point to epsilon greater than the smallest of its neighbors and then recalculating flows.  This procedure leads to the creation of cycles of 3 or 4 points.  For instance, consider A and B flowing into C, where C is a local minimum.  Increasing the height of C causes it to flow into A.  If A now is a local minimum, it will flow into B after filling, creating a cycle.

For small enough values of m, however, our model appropriately returns a ``swamp" of many short basins terminating in lakes.   The Delaunay triangulated random grid, by adding noise in the locations of points and thus the difference in height between neighbors, effectively added randomness to the system without creating undue lakes.   

The algorithm to find magnitude and Strahler numbers for each lattice point and link in the network is recursive by necessity.  Additionally, the matrix of connectivities is $n^2$x$n^2$.  We were thus limited in the size of the networks we could simulate to approximately $r=67$.  For the purposes of finite size scaling analysis (see below), we simulated networks at $r=100$, however these simulations took several hours each, due to paging.

\section{Finite Size Scaling}

\begin{figure*}
\includegraphics{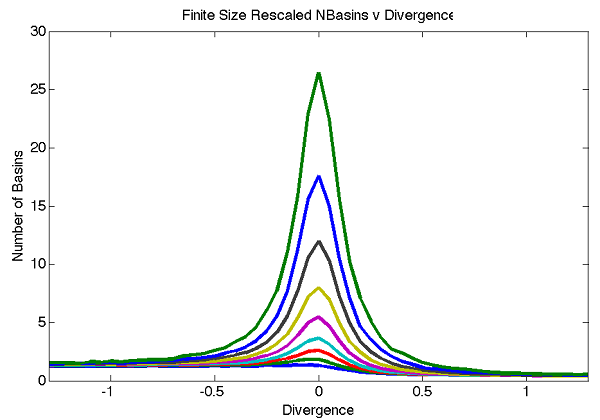}
\caption{(Color Online) Finite Scaling analysis:  We plot the reduced number of basins, $N_{basins}/r^\Lambda$, where $\Lambda=1$, the finite size scaling exponent for logarithmically spaced values of $r$ from 4 (at the bottom) to 100 (at the top)}
\end{figure*}

We conducted simulations for $r=4$ to $r=100$ (the different colored lines) at logarithmically-spaced intervals (a factor of 1.5, rounded down) in the range $m=-1.3$ to $m=1.3$ stepping by 0.05.  We then logarithmically regressed versus $r$ for each point with $|m|>1$.  This gave us a finite-size scaling exponent $\Lambda=1\pm0.04$. The results of $N_{basins}/r^\Lambda$ vs. $m$ for each value of $r$ are plotted in Figure S1.  We then logarithmically regressed the rescaled $N_{basins}$ with $r$ to find the critical exponent, $\gamma$.  This procedure was repeated for $\bar{l}$, to find $\delta$.  

\section{Additional Results}
The original, planar, Scheidegger Model is not recovered at  $m=0$.  The model we simulate lacks a uniform slope in either the x or y direction, which is implicit in the original Scheidegger model.  This uniform slope could be included by altering our slope field to read: $z=mr+ny+\eta$.  Geometrically, this modification would tilt the conical manifold we simulate towards either the positive or negative y-direction.  One may envision a tilted coffee filter that is flattening as m approaches zero.  In essence, it would add an additional ``field" which would split and shift the location of the critical point to $m=\pm n$.  For$\|m\|> n$, the model would function as before.  At the first critical point crossed, half the manifold would be nearly flat and would exhibit a singularity in number of basins, whereas the other half would be tilted at the sum of the two slopes.  As m approached zero, we would recover the original Scheidegger model.  At the second critical point, the opposite half of the manifold would be nearly flat and would exhibit the same singularity.  The total effect would be akin to the addition of an external field to the ferromagnetic phase transition.

%\bibliography{PRLNetworksBib}
%
\end{document}